\documentclass [10pt]{article}
\usepackage {amssymb}
\usepackage {amsmath}
\usepackage [cp1251]{inputenc}
\usepackage{graphicx}
\usepackage {longtable}

\begin{document}
\begin{center}
\textbf{Geometry and dynamics of billiards in symmetric phase space}
\end{center}

\bigskip

\begin{center}
S.V. Naydenov \footnote{E-mail: naydenov@isc.kharkov.com}, V.V.
Yanovsky
\end{center}

\bigskip

\begin{center}
Institute for Single Crystals of National Academy of Sciences of Ukraine,
\end{center}

\begin{center}
60 Lenin ave., 61001 Kharkov, Ukraine
\end{center}

\bigskip

\begin{center}
\textbf{Abstract}
\end{center}

The billiard problem of statistical physics is considered in a new geometric
approach with a symmetric phase space. The structure and topological
features of typical billiard phase portrait are defined. The connection
between geometric, dynamic and statistic properties of smooth billiard is
established. Other directions of the theory on development are pointed out.

\bigskip

\begin{center}
\textbf{1. Introduction}
\end{center}

Billiard is one of the most important models of statistical physics and
chaotic dynamics. G.D. Birkhov suggested regarding billiard as a typical
conservative system [1]. A.N. Krylov based his explanation of solid spheres
gas statistic properties on exponential divergence of its ``billiard''
trajectories [2]. In the works by Ya.G. Sinai and L.A. Bunimovich on phase
trajectories mixing in scattering and defocusing billiards Boltzmann's
hypothesis of molecular chaos found its further grounding [3]. Now billiard
became a paradigm of deterministic chaos [4] of classical systems and is
often applied [5] for the research of their quantum ``twins''. A lot of
applied physics problems can be reduced to a billiard problem [6].

A classical billiard problem is in studying its character and
distribution of its trajectories. Among the typical billiard
motions one can point out the following: periodical,
quasiperiodical (integrable) and irregular (chaotic) motions.
Compound billiard dynamics appears in the phase portrait structure
of the corresponding mapping. The latter is plotted using
different geometric methods or Poincare sections. For the
specification of a billiard ray it's usual to choose local Birkhov
coordinates: natural parameter $l$ in the reflection point on the
border of the billiard $\partial \;\Omega $ and the incidence
angle $\theta $ in the same point. They stand for canonical
variables -- the coordinate and the moment for Hamiltonian
description of the system. Many important properties at this
choice of phase space coordinates stay unnoticed. Let us choose
another unifying approach. It identifies billiards with reversible
mapping (with projective involution) in a symmetrical phase space.
In its framework one can join together geometric, dynamic and
statistic properties of billiards. Such fundamental mechanisms are
analyzed in this paper.

\bigskip

\begin{center}
\textbf{2. Symmetric Coordinates}
\end{center}

Let us describe geometric propagation of the rays of billiard (Fig.) as a
reversible mapping $B$ of the phase space $Z_{{\rm} }$with symmetric
coordinates $(z_{1} ,z_{2} )$. The pair of these coordinates defines two
successive reflections of a billiard ray from $\partial \;\Omega $. At the
same time, each of the coordinates corresponds to some parameterization of
the billiard border, ${\left. {\vec {r}} \right|}_{\partial \Omega}  = \vec
{r}(z) = (x(z),y(z))$. The following topological construction appears: $Z
\propto \partial \;\Omega \times \partial \;\Omega $. For a closed planar
billiard one can accept $z \in S^{1}$ (circle) or $z \in I = [0,1]$ the
periodicity being $\vec {r}(z) = \vec {r}(z + 1)$. So we'll have a phase
space as a torus $Z = T^{2} = S^{1}\times S^{1}$ or its unfolding $\Pi =
I\times I$ on the plane. After each reflection of an arbitrary (incoming)
billiard ray with the coordinates $(z_{1} ,z_{2} )$, we have a (reflected)
ray with new coordinates $({z}'_{1} ,{z}'_{2} )$. As a result, the evolution
of these successive reflections is described with a billiard cascade $(z_{1}
,z_{2} ) \to ({z}'_{1} ,{z}'_{2} )$ of the form [7]

\begin{equation}
\label{eq1}
B: = {\left\{ {{\begin{array}{*{20}c}
 {{z}'_{1} = z_{2}}  \hfill \\
 {{z}'_{2} = f(z_{1} ,z_{2} )} \hfill \\
\end{array}} ;{\kern 1pt}}  \right.}f(f(z_{1} ,z_{2} ),z_{2} ) = z_{1} {\rm
;}
g(f,z_{2} ) = g(R(z_{1} ,z_{2} ),z_{2} ));\;R(z,{z}') = {\frac{{a({z}')z +
b({z}')}}{{b({z}')z - a({z}')}}}
\end{equation}

\noindent with the involution $f = f(z_{1} ,z_{2} )$ (on the first
argument$z_{1} $), which is defined by implicit dependence on the
corresponding fractional rational involution $R$,
$R(R(z,{z}'),{z}') = z$. The coefficients $a(z) = n_{x}^{2} (z) -
n_{y}^{2} (z)$; $b(z) = 2n_{x}^{2} (z)n_{y}^{2} (z)$ are expressed
with (Cartesian) components of the exterior normal field $\vec
{n}(z) = \vec {n}_{ext} {\left| {_{\partial \Omega} }  \right.}$
on the border $\partial \;\Omega $, $\vec {n} = (n_{x} ;n_{y} ) =
({y}'(z); - {x}'(z))$ (the stroke marks differentiation). Function
$g$ depends on the form of $\partial \;\Omega $ and $g(z_{1},z_{2}
) = g(z_{2} ,z_{1} )=$ $={\left[ {x(z_{1} ) - x(z_{2} )} \right]}
/ {\left[ {y(z_{1} ) - y(z_{2} )} \right]}$. The choice of
Cartesian coordinates is rather relative, because every covariant
substitution $z_{1} \to \xi B(z_{1} );z_{2} \to \xi B(z_{2} )$
preserves the form of the mapping (\ref{eq1}). The mapping built
can be used for the description of billiard with a border of the
most general type. During its derivation only the condition of
elastic reflection and no restrictions of smoothness, curvature,
convexity, simple connectivity of the border and so on were
introduced.

The mapping (\ref{eq1}) is invariant to the substitution, $S: = \;z_{1} \to z_{2}
;z_{2} \to z_{1} $, of the incoming ray to the reflected one, i.e. $B \circ
S = S \circ B$ for the composition of transformations. This means
reversibility of the constructed mappings. The physical reason of this is
reversibility of the system to the changes of the time sign (the direction
of the motion). This is a global property. In the billiard cascade, phase
trajectories with opposite directions of the motion or with
opposite-directional initial rays $(z_{10} ,z_{20} )$ \`{e} $(z_{20} ,z_{10}
)$are present simultaneously. This requirement of local reversibility is
stronger. The inverse of the ray reflected with its successive reflection
makes the initial incoming ray. Mathematically it leads to the appearance of
a involution $f$ in the mapping (\ref{eq1}). The symmetry (reversibility) leads to
the symmetry of the phase space and the phase portrait of the mappings (\ref{eq1}).
For every element $Z$ there is one symmetrical to it relative to the
diagonal $\Delta = {\left\{ {(z_{1} ,z_{2} ) \in Z\,{\left| {\;z_{1} = z_{2}
} \right.}} \right\}}$. For every function $\chi $ on $Z_{{\rm} }$

\begin{equation}
\label{eq2}
\chi (z_{1} ,z_{2} ) = \chi (z_{2} ,z_{1} ) \quad {\rm .}
\end{equation}

That's why it's natural to regard the coordinates of $Z$ as symmetrical.
This symmetricity essentially sets them apart from the variables of
Hamiltionian description of the billiard. In particular, the locality of
Birkhov coordinates causes explicit mappings to be obtained only for the
simplest geometry $\partial \;\Omega $. In non-local coordinates, $(z_{1}
,z_{2} )$ this problem can have a general solution.

In the research of periodical trajectories of the billiard the powers of
billiard mapping $B^{k}$ are also used

\begin{equation}
\label{eq3}
B^{k}: = {\left\{ {{z}'_{1} = f_{k - 1} (z,z);{z}'_{2} = f_{k} (z_{1} ,z_{2}
)} \right\}}{\rm ;}
\quad
f_{k} (z_{1} ,z_{2} ) = f(f_{k - 2} (z_{1} ,z_{2} ),f_{k - 1} (z_{1} ,z_{2}
)) \quad {\rm .}
\end{equation}

They include billiard ``compositions'' $f_{k} $, where $k = 0,1,2,\ldots $;
$f_{1} = f$; $f_{0} = z_{2} $; $f_{ - 1} = z_{1} $. They lose the property
of involution, but preserve reductibility to fractional rational
transformations. The mappings (\ref{eq3}) describe ``pruned'' billiard trajectories
with the omission of a set of $(k - 1)$ links (successive reflections).

\begin{center}
\textbf{3. Billiard Geometry: Involution Properties}
\end{center}

All the geometric properties of a billiard are established in the
specialization of mappings (\ref{eq1}). They are concretized in the features of the
involution$f$. In the appropriate (local) coordinates it can be reduced to
fractional rational involution $R$. Projective transformations are described
with fractional rational functions. Billiard is one of those
transformations. In every reflection point incoming and reflected rays are
joined together by a harmonic transformation $G$. For $G$ projective
invariant (a complex relation of four rays , incoming $i$, reflected $r$,
normal $n$ and tangent $t$) is equal to $(i,r,n,t) = - 1$. In geometric
terms involution looks the simplest

\begin{equation}
\label{eq4}
r = G(i;n,t){\rm ;}
\quad
G \circ G = Id \quad {\rm ,}
\end{equation}

$Id$ is an identical transformation. Let us emphasize the locality
of the projective property of the billiard. The concrete form of
$G$ depends on the guiding-lines of the normal in the point of
reflection, that is, on the form $\partial \;\Omega $. Harmonic
mapping (\ref{eq4}) is an involution and changes the sequence
order of the ordered projective elements to the opposite. The
monotony of $f$ on the first argument is the consequence of this.
This monotony of piece-wise continuous $f$ (involution can stand
discontinuity) is true for every billiard. Using the correlations
(\ref{eq1}), the involution can be laced of local branches of the
form $f = g^{ - 1} \circ R \circ g$ (the choice of $g^{ - 1}$
branch is dictated by the variational principle, the minimality of
distance between the points of reflection). From that the property
of monotony immediately follows

\begin{equation}
\label{eq5}
\partial {\kern 1pt} f(z_{1} ,z_{2} ) / \partial {\kern 1pt} z_{1} < 0 \quad
{\rm .}
\end{equation}

Fractional rational functions are dense everywhere in the space of
continuous functions. In fact, this means the possibility of
arbitrary precise approximation of different physical systems with
their billiard models. This fact is used, for instance, in the
analysis of energetic spectra of multi-particle systems, the
description of kinetic properties of continuum (Lorenz gas model)
etc. If sinus and cosine have physically appeared from the problem
of oscillator, then fractional rational functions can be generated
by billiard.

The reflection of ray beams from the border of the billiard can be of
diffractive, focusing and neutral character. This depends on the curvature
of $\partial \;\Omega $. The representation (\ref{eq1}) gives the following property

\begin{equation}
\label{eq6}
sign\;{\left\{ {{\frac{{\partial {\kern 1pt} f(z_{1} ,z_{2} )}}{{\partial
{\kern 1pt} z_{2}} }}} \right\}} = sign\;{\left\{ {\hat {K}(z_{2} )}
\right\}} \quad {\rm ,}
\end{equation}

\noindent where $\hat {K}$ is oriented curvature in the point of
reflection. For the convex border $\hat {K} > 0$ involution
appears to be a monotonous function on both arguments. (For
instance, for a circle, $f(z_{1} ,z_{2} ) = 2z_{2} - z_{1} \quad
(\bmod \;1)$.) On a torus, $Z = T^{2}$, such involution has no
breaks (Unlike everywhere dispersive billiard, $\hat {K} < 0$,
with lacunas in phase space.)

Involutivity and projectivity are the main geometric properties of a
billiard. The geometry (form) of its border defines the explicit form of
involution. At the same time, it also defines the dynamics of the billiard.

\begin{center}
\textbf{4. Billiard Dynamics: the Structure of Symmetric Phase
Space}
\end{center}

Let us analyze the structure of symmetric phase space of a typical billiard
(see Fig.). This principally solves the question of the types of dynamics
and stability. For more simplicity let us regard the border $\partial
\;\Omega $ as a differentiable as many times as needed. The research of
billiards in polygons (without curvature) needs some peculiarities.

For high-quality research of phase portrait of the mappings and
its local bifurcations normal Poincare forms are especially useful
[8]. In the symmetric approach the theory of normal billiard forms
appears to be the most advanced. This is connected with the
flexibility (a wider class of allowable variables) of reversible
systems. Any changes of variables in Hamiltonian approach are to
preserve the conservation character of the mapping with the
Jacobian $J = 1$ (canonical changes). Whereas the mapping
(\ref{eq1}) doesn't demand it. It Jacobian $J = - \partial {\kern
1pt} f(z_{1} ,z_{2} ) / \partial {\kern 1pt} z_{1} > 0$ can take
arbitrary values, $0 < J \le 1;\;J \ge 1$. As a weak limitation,
the demand for the mapping (\ref{eq1}) to preserve measure
remains. This means that $J = J(\vec {z}) = \rho (\vec {z}) / \rho
(B(\vec {z}))$, where $\vec {z} = (z_{1} ,z_{2} );\;B\,\vec {z} =
(z_{2} ,f(z_{1} ,z_{2} ))$ should be true. The proof uses the
equation of Frobenius--Perron for the density $\rho $ of invariant
measure (see further) and the symmetry (\ref{eq2}) for it, $\rho
(z_{1} ,z_{2} ) = \rho (z_{2} ,z_{1} )$. This limitation can
always be met preserving the main property of involution $f \circ
f = id$ in new coordinates.

Omitting the details, let us present the expression for universal normal
billiard form in symmetric coordinates. It is true in the neighbourhood of
an arbitrary cycle of $p$order (periodic trajectory of $p$period)

\begin{equation}
\label{eq7}
NB^{p}: = {\left\{ {{\begin{array}{*{20}c}
 {{z}'_{1} = - \mu _{p - 1} z_{1} + \nu _{p - 1} z_{2} + z_{1} P(z_{1} z_{2}
)} \hfill \\
 {{z}'_{2} = - \mu _{p} z_{1} + \nu _{p} z_{2} + z_{2} Q(z_{1} z_{2} )}
\hfill \\
\end{array}} } \right.} \quad {\rm ,}
\end{equation}

\noindent
where the coefficients of the linear part are defined by the expansion of
``compositions'' $f_{p - 1} $ and $f_{p} $ (see formula (\ref{eq3})) in the initial
point neighbourhood of the cycle under consideration. They constitute the
matrix of $\hat {L}$ linear part. Its determinant is equal to one, $\det
\,\hat {L} = 1$, that agrees with conservation of measure and the condition
$J(C) = 1$ for every cycle $C$ of mapping (\ref{eq1}). Homogeneous polynomials $P,Q$
(without absolute terms) define nonlinear additives. Their explicit form
depends on the involution of billiard $f$, that is, on the form of $\partial
\;\Omega $.

The character of the cycle depends on the size of trace $tr\hat {L}$. For an
elliptic cycle ${\left| {tr\;\hat {L}} \right|} < 2$, for a hyperbolic one
${\left| {tr\;\hat {L}} \right|} > 2$. In the neutral case, for instance,
for a billiard in a circle, $tr\hat {L} = 2$. It can be shown that for any
cycle, corresponding to a periodic trajectory, passing through a concave
section with concavity $\hat {K}(z_{2} ) < 0$, $tr\hat {L} < - 2$ will be
true. That's why the trajectories near such cycles always are unstable and
exponentially diverge from one another. Near elliptic cycles, including
2-cycles, regions of regular motion form. With the loss of ellipticity they
are ruined, first forming stochastic layers and then, when the latter are
covered, a chaotic sea. In reality this mechanism looks much more
complicated, for example, with an intermediate pass through Cantor-tori etc.
Normal forms (\ref{eq7}) let us trace typical properties of such bifurcations,
taking place when the billiard border is deformed.

The diagonal $\Delta (z_{1} = z_{2} )$ contains all fixed points of the
billiard, $B\Delta = \Delta $. This follows from the diagonal property
$f(z,z) = z$ of billiard involution, resulting from its coordinate
expression (\ref{eq1}). For a convex billiard in the neighbourhood of the phase
space diagonal, normal form (\ref{eq7}) can be reduced to the mapping of a turn,
i.e. a particular case of a billiard in a circle. Here the structure of
elliptic and hyperbolic cycles of arbitrary high order is shown. The motion
stays regular. The appearing of negative curvature ruins this situation.
There is no unified transformation (or the integral of motion) near the
diagonal because of appearing breaks of billiard involution.

Analytic research of the symmetric phase space structure can be continued
using geometric methods. Here the important advantages of the new approach
are seen. In addition to regular and chaotic components of motion the phase
portrait can contain regions of forbidden motion -- ``lacunas'' $L_{{\rm
}}$and regions of degenerated motion -- ``discriminants'' $D$.

Lacunas (Fig.) appear in the billiards with regions of negative curvature.
They occupy the phase space part, the points of which correspond to the rays
lying outside of the billiard region $\Omega $. The coordinates of these
rays meet the condition $\vec {r}(z_{1} ) - \vec {r}(z_{2} ) \notin \Omega
$. This condition defines the inner region of lacuna $L$ in $Z$. The form of
the lacuna is defined by its border $\partial \,L: = \{(z_{1} ,z_{2} ) \in
Z{\left| {} \right.}z_{1} = \lambda (z_{2} )\}$. (Another parameterization
is possible $z_{2} = \lambda ^{ - 1}(z_{1} )$. In this case functions
$\lambda (z)$ and $\lambda ^{ - 1}(z)$ specify the same simple closed curve,
but passable in different directions. When one its branch is higher than
$\Delta $and the other is lower, and vice versa.) The border $\partial \,L$
comes to the diagonal $\Delta $ transversally and crosses it twice in the
points with coordinates $(z_{0} ,z_{0} )$, corresponding to the points of
inflexion, $\hat {K}(z_{0} ) = 0$.

The forbidden billiard rays (points of lacuna) lie in classically
inaccessible region -- geometric shade, generated by the regions $\hat {K} <
0$. The number of lacunas (on a torus $Z = T^{2}$ ) is equal to the number
of negative curvature components $\partial \;\Omega $. Every lacuna is a
simply connected set. The contrary would mean non-closed character of
$\partial \;\Omega $. With the appearance of lacunas a part of diagonal
$\Delta $ is cut out. The corresponding fixed points disappear. For
everywhere dispersive Sinai billiard, lacuna absorbs all the diagonal and
the mapping (\ref{eq1}) will lack all fixed points. At a special configuration of
such a border $\partial \;\Omega $ one can cut out cycles of higher order,
$p \ge 2$.

In a topological way one can glue up the lacuna on the torus with a
two-dimensional manifold. According to the rule of $\partial \;L$ bypass, it
can be only a piece of a projective plane. This is directly connected with
the projectivity of the billiard. On a projective plane, metric conceptions
``inside'' and ``outside'' of a closed region lose their sense. (For
example, a closed curve and a right line that doesn't cross it on a plane
may have common points after central projection onto the other plane.)
That's why ``forbidden'' rays turn out to be involved into the general
billiard flow. Such global motion takes place on non-oriented manifold.

On the projective plane the initial involution $f$ also rules the motion of
the rays. Almost every such ray (exceptional cases are of measure null) is
continued to an ordinary billiard ray, further dynamics of which is known.
As a result of further evolution, this ray after some time will return to
the section of negative curvature $\partial \;\Omega $, corresponding to the
lacuna under consideration. This is specified by the mentioned hyperbolicity
of cycles that contain points on the concave border. Being continued then to
a classically inaccessible region (preserving the direction of motion), it
would give a new position of the initial ray (phase point in the lacuna). A
recurrent mapping appears. It is defines by one of the ``compositions''
$f_{k} $, included in the equation (\ref{eq3}), the order $k$ always depending on
the coordinates of the initial ray (the initial point of the lacuna). The
lacuna plays the role of a secant for the Poincar\'{e} section of the
billiard flow. Similar evolution also takes place with other points of all
lacunas. Only phase trajectories of ordinary billiard rays remain in this
case ``visible''.

The condition of ``connecting'' for the inner rays, that are tangent to the
concave region in the point $\vec {r}(z_{3} )$ and that cross $\partial
\;\Omega $ in the points $\vec {r}(z_{1} )\,,\,\vec {r}(z_{2} )$ outside of
it, defines the border of the corresponding lacuna

\begin{equation}
\label{eq8}
\left( {\vec {r}(z_{1} ) - \vec {r}(z_{2} ),\;\vec {n}(z_{3} )} \right) =
0\;;\;(z_{1} ,z_{2} ) \in Z\vert {\kern 1pt} L\,;\;(z_{1} ,z_{3} ) \in
\partial \,L\;;\;(z_{3} ,z_{2} ) \in \partial \,L \quad {\rm ;}
\quad
\hat {K}(z_{3} ) < 0 \quad {\rm ,}
\end{equation}

\noindent
where $(.,.)$ is the scalar product of the vectors. Solving the equation (\ref{eq8})
according to the theorem about an implicit function, we have $z_{1} =
\lambda _{1} (z_{3} );\;z_{2} = \lambda _{2} (z_{3} )$. Excluding $z_{3} $,
we come to the desired equation $z_{1} = \lambda (z_{2} );\;\lambda =
\lambda _{1} \circ \lambda _{2}^{ - 1} $.

The discriminants $D$ correspond to the zone of ``stuck-together''
trajectories or ``non-continuable'' trajectories that cross special (corner)
points of $\partial \;\Omega $. That's why they appear in the billiards with
straight regions of borders, $\hat {K} = 0$. Topologically the discriminants
have a lacuna-like structure. Their border $\partial \;D$ is defined by

\begin{equation}
\label{eq9}
\left( {\vec {r}(z_{1} ) - \vec {r}(z_{2} ),\;\vec {n}(z_{2} )} \right) =
0\,;\;(z_{1} ,z_{2} ) \in Z\vert L \quad {\rm ;}
\quad
\hat {K}(z_{2} ) = 0 \quad {\rm .}
\end{equation}

It also can have explicit form $z_{1} = \mu (z_{2} )$. The discriminants
have regular shape -- squares (Fig.) with a diagonal, which coincide with a
part of $\Delta (z_{1} = z_{2} )$ in the region corresponding to the
straight-line component $\partial \;\Omega $.

Lacunas and discriminants make a principal property of a symmetric phase
space. In fact, they are filled with the rays of the billiard that fell out
of its ordinary dynamics. (At their passing it's easy to show that the
billiard involution $f$ breaks (on the first and the second arguments), the
breaks being different from the factor of periodicity $(\bmod \,1)$ and are
not removed when passing to a torus, $Z = I^{2} \to Z = T^{2}$.) There are
no such non-local elements in the phase space of Hamiltonian approach. At
the same time, these formally hidden ``topological'' obstacles (Fig.) for
the billiard flow to flow around, and the diagonal $\Delta $, on which they
arise, play an important role in the chaotic dynamics and must be included
into the full description.

\begin{center}
\textbf{5. Billiard Kinetics: Invariant Distributions}
\end{center}

The geometry of phase space structural elements depends on the form of the
border $\partial \;\Omega $ and (or) involution $f$. Let us show that in the
symmetrical approach not only dynamics but also kinetics of the billiard is
connected with these characteristics. The kinetics becomes apparent in the
case of chaotic billiard, whose deterministic trajectories (unique groups of
successive points of reflection) have all the properties of random sequences
in the asymptotic limit of infinitely large number of reflections. That
requires statistic description of (two-dimensional) dynamic system in the
manner of deterministic chaos conception [9].

In the mixed dynamics of a typical billiard both integrable and ergodic (as
a rule, with intermixing) types of motion are present. In this case, the
general characteristic of the trajectories distribution in the billiard is
the invariant measure of the dynamic system. Absolutely continuous
distributions are of the greatest physical interest. It's important for them
to be individualized, i.e. depending on the geometry of a concrete billiard.
That's why universal measures like conserved Liouville volume (Birkhov
measure [1] for all billiards) are of little use here. A symmetric measure
with density $\rho \left( {z_{1} ,z_{2}}  \right) = \rho \left( {z_{2}
,z_{1}}  \right)$ already possesses this individuality.

From the operator equation $B\rho = \rho $ for an invariant measure after
transformations using piece-wise monotony (\ref{eq5}) we have

\begin{equation}\label{eq10}
\rho (z_{1} ,z_{2}) = \rho (z_{2} ,z_{1} ) = \rho
(z_{2},f(z_{1},z_{2})) \left( -\frac { \partial {\kern 1pt}
f(z_{1},z_{2}) }{ \partial z_{1} } \right) = \rho (z_{2},f(z_{1}
,z_{2})) J(z_{1},z_{2}) \quad {\rm .}
\end{equation}

In some cases (integrable billiards) one can find its explicit solution.
Geometrically, $\rho $ is a two-point density; it depends on the coordinates
of two points on the border $\partial \;\Omega $. The topology of the direct
product $Z \propto \partial {\kern 1pt} \Omega \times \partial {\kern 1pt}
\Omega $ causes one to choose a special factorized solution, $\rho (z_{1}
,z_{2} ) = \omega (z_{1} )\,\omega (z_{2} )$. Instead of the expression (\ref{eq10})
we get a functional equation for one-point plane $\omega (z)$:

\begin{equation}
\label{eq11}
\omega \,\,(z) = \omega \,(f)\,J(z,{z}');\;f = f(z,{z}')\quad
\Leftrightarrow \quad \omega \,(z)\,dz = - \omega \,(f)\,df \quad {\rm ,}
\end{equation}

\noindent written in total differentials. The factorization is
coordinated with the symmetry of $\rho $ and preserves its
normalization ${\left\| {\omega} \right\|} = {\int_{0}^{1} {\omega
(z){\kern 1pt} dz = 1}} $.

The physical sense of $\omega (z)$ is an asymptotic plane of billiard flow
reflection points (with coordinates $\vec {r}(z) \in \partial {\kern 1pt}
\Omega ,\quad z \in I$). This is a truncate distribution in the sense that
the dimension falls twice. It will be very useful in the description of
physical characteristics in different billiard problems, for instance, the
``probability'' of ray escaping from a fixed place of resonator, wave-guide
or detector. Besides, it is directly connected with the involution and
geometry of the billiard. After integrating the differential relation (\ref{eq11})
for $\omega $ we have

\begin{equation}
\label{eq12}
\left( {{\int\limits_{z_{0}} ^{f} { - {\int\limits_{z_{1}} ^{z_{0}}  {}} }
}} \right)\omega {\kern 1pt} (z){\kern 1pt} dz = C(z_{2} )\,;\quad f =
f(z_{1} ,z_{2} ) \quad {\rm ,}
\end{equation}

\noindent
where $z_{0} $ is an arbitrary initial point on $\partial \;\Omega $; $C(z)$
is the function to define. With different character of border $\partial
\;\Omega _{{\rm} }C(z)$ has different forms. For everywhere convex
billiard it's one can just use the diagonal condition $f(z,z) = z$, so $C(z)
= 2{\int_{z_{0}} ^{z} {\omega {\kern 1pt} ({z}')\,d{z}'}} $. In the general
case the border $\partial \;\Omega = \partial \,\Omega _{ +}  \cup \partial
\,\Omega _{ -}  \cup \partial \,\Omega _{0} $ contains regions of positive,
$\partial \;\Omega _{ +}  $, negative, $\partial \;\Omega _{\_} $, and zero,
$\partial \;\Omega _{0} $, curvature. During the defining of $C(z)$ the
solutions in symmetric ``halves'' of phase space over and under the diagonal
$\Delta $, that is, in the involutionally connected regions with coordinates
$(z_{1} ,z_{2} )$ and $(f(z_{1} ,z_{2} ),z_{2} )$ are laced. In the presence
of $\partial \;\Omega _{\_} $ and $\partial \;\Omega _{0} $ components
connecting takes place on the borders of corresponding lacunas and
discriminants. Summing it up, let us set the border $\partial \;\Sigma $,
that divides different symmetric components of ordinary cascade $\Sigma $
(outside special zones)

\begin{equation}
\label{eq13}
(z_{1} ,z_{2} ) \in \partial \,\Sigma \quad \Leftrightarrow \quad z_{1} =
\Lambda (z_{2} ) = {\left\{ {{\begin{array}{*{20}c}
 {z_{2} \,,\quad (z_{1} ,z_{2} ) \in \Delta}  \hfill \\
 {\lambda (z_{2} ),\quad (z_{1} ,z_{2} ) \in \partial \,L} \hfill \\
 {\mu (z_{2} ),\quad (z_{1} ,z_{2} ) \in \partial \,D} \hfill \\
\end{array}} } \right.}
\end{equation}

\noindent
with known dependencies in the cases of lacunas and discriminants (see
above). Let us note that in each half of the phase space $\Lambda (z)$ is a
multi-valued function (the number of branches doesn't exceed the doubled
number of $\partial \;\Omega _{\_} $ and $\partial \;\Omega _{0} $
components, but self-intersections and multiple connection $\partial
\;\Sigma $ are forbidden by the uniqueness of the flow. $\Lambda (z)$ is the
functional of $\partial \;\Omega $form. Connecting on the border $\partial
\;\Sigma $ gives us

\begin{equation}
\label{eq14}
\left( {{\int\limits_{z_{0}} ^{f(\Lambda (z),\,z)} { - {\int\limits_{\Lambda
(z)}^{z_{0}}  {}} }} } \right)\omega {\kern 1pt} ({z}'){\kern 1pt} {\kern
1pt} d{z}' = C(z)\quad \Rightarrow \quad
\left( {{\int\limits_{\Lambda (z_{2} )}^{f(z_{1} ,{\kern 1pt} z_{2} )} { -
{\int\limits_{z_{1}} ^{f(\Lambda (z_{2} ),{\kern 1pt} z_{2} )} {}} }} }
\right)\omega {\kern 1pt} (z){\kern 1pt} dz = 0 \quad {\rm .}
\end{equation}

The dependence of the initial point $z_{0} $, as would be expected, falls
out. The equation obtained lets one to restore billiard involution $f$ on
the one-point billiard distribution function $\omega _{{\rm} }$and vice
versa. At the same time both functions are connected with the equation of
border $\partial \;\Omega $ by the expressions (\ref{eq13}) and (\ref{eq1}). The billiard
problem takes on a single meaning from dynamic, statistic and geometric
points of view.

Direct solution for $\omega $ on $f$ can be obtained by
differentiation of Eq.(\ref{eq11})

\begin{equation}
\label{eq15}
{\frac{{d\ln \omega {\kern 1pt} \,(f)}}{{d{\kern 1pt} f}}} = -
{\textstyle{{{\raise0.7ex\hbox{${\partial ^{2}f(z_{1} ,z_{2} )}$}
\!\mathord{\left/ {\vphantom {{\partial ^{2}f(z_{1} ,z_{2} )} {\partial
z_{1} \partial {\kern 1pt} z_{2}
}}}\right.\kern-\nulldelimiterspace}\!\lower0.7ex\hbox{${\partial z_{1}
\partial {\kern 1pt} z_{2}} $}}} \over {{\raise0.7ex\hbox{${\partial {\kern
1pt} f(z_{1} ,z_{2} )}$} \!\mathord{\left/ {\vphantom {{\partial {\kern 1pt}
f(z_{1} ,z_{2} )} {\partial
z}}}\right.\kern-\nulldelimiterspace}\!\lower0.7ex\hbox{${\partial z}$}}_{1}
\;{\raise0.7ex\hbox{${\partial {\kern 1pt} f(z_{1} ,z_{2} )}$}
\!\mathord{\left/ {\vphantom {{\partial {\kern 1pt} f(z_{1} ,z_{2} )}
{\partial {\kern 1pt} z_{2}
}}}\right.\kern-\nulldelimiterspace}\!\lower0.7ex\hbox{${\partial {\kern
1pt} z_{2}} $}}}}} \quad {\rm .}
\end{equation}

Having performed the change $z_{1} \to f(z_{1} ,z_{2} )$ and using
the relation $f \circ f = id$ for calculating the derivatives,
this equation for $\omega $ can be written as

\begin{equation}
\label{eq16} {\left[ \ln \omega (z_{1})\right]}^{\prime} _{z_{1}}
= - \frac{f''_{z_1 z_2} }{ {f'_{z_1}}^{3} \, {f}'_{z_2} } + \frac{
{f''}_{z_z^2} }{ {f'_{z_1}}^4 } ;\; f = f(z_{1},z_{2}) \quad ,
\end{equation}

\noindent
where strokes mark partial derivatives. The right part of the equation (\ref{eq16})
shouldn't depend on $z_{2} $. This limits, on the one hand, the type of
involution (not every involution can be billiard, that corresponds to the
stressed role of projective transformations), and, on the other hand, the
choice of phase space variables, allowing factorization of two-point measure
to one-point ones. Covariant changes, leading to ``factorizing''
coordinates, correspond to the choice of a certain frame and way of border
$\partial \;\Omega $ parameterization.

In the equations (\ref{eq10}) and (\ref{eq16}) the densities $\rho $ and $\omega $ are
uniquely defined by the involution of billiard $f$. The latter is uniquely
defined by the border $\partial \;\Omega $equation, according to the
representation (\ref{eq2}). The invariant measures of the billiard become its
individual characteristics. In a chaotic billiard they acquire the character
of equilibrium statistic distributions.

So, on the whole billiard analysis in symmetrical coordinates shows that its
main characteristics are uniquely connected with one another

\begin{equation}
\label{eq17}
\partial \,\Omega \to f(z_{1} ,z_{2} ) \to \rho (z_{1} ,z_{2} )
\leftrightarrow \omega \,(z)\; \to f(z_{1} ,z_{2} ) \to \partial \,\Omega
\end{equation}

\noindent
where arrows indicate passing from one object to another.

One of the most designing and old problems of statistical physics is finding
out the transition from the reversibility of deterministic motion equations
to irreversibility of statistic ones, see for ex. [10]. Generally accepted
point of view is that irreversibility appears at roughening in the
macroscopic description of the system on the kinetic stage of evolution and
is connected with the fundamental principle of correlations unlinking. Here
usually the problem of distribution functions calculation with given
Hamiltonian (the equations of motion) is posed. In physical applications the
inverse problem may also appear: to restore the dynamic law for a chaotic
system (not necessarily of mechanic origin) with known statistic
characteristics. It can be of special actuality for the system with a small
number of freedom degrees.

Statistic irreversibility prevents the reverse of the ``time arrow'', but
doesn't necessarily break the feedback of kinetic and dynamic. A remarkable
peculiarity of the billiard is the possibility to solve direct as well as
indirect problems. The form of the border $\partial \;\Omega $ defines
involution, on which the invariant measure is calculated. And vice versa:
the involution (that is, the dynamic of the billiard) is restored from the
one-point distribution of reflections on the border. The border of the
billiard can be restored by its involution [7]. By the way such closure is
the consequence of geometric (projective) nature of the billiard.

\begin{center}
\textbf{6. Summary}
\end{center}

In the conclusion let us note the characteristic properties of the
symmetrical approach.

1. The symmetricity of the phase space. Equality of phase
coordinates rights. Non-local character of the geometric elements
involved into the dynamics.

2. Unification of billiards as reversible dynamic systems
(mappings) with projective involution. Reversibility and
projectivity of the billiard.

3. Geometric character of the phase space structure, taking into
account billiard border properties (its diagonal, lacunas,
discriminants). Universal character of symmetrical normal forms,
describing the dynamics and local bifurcations in the
neighbourhood of cycles and special zones of the billiard.

4. Individualization of invariant distributions defined by the
form of the billiard. The reduction of the measure (one-point
factorization) without loss of statistic description. The division
of quick and slow variables is not necessary.

5. The unity of dynamics, kinetics and geometry of the billiard.
The solution of direct and indirect problems on the restoration of
involution, invariant measure and the form of the border.

Symmetric approach allows direct generalization on the multiply
connected, multi-dimensional and other cases of different billiard
border topology. The peculiarities named preserve here their key
role.

\bigskip

\begin{center}
\textbf{Acknowledgment}
\end{center}

The authors feel great pleasure to express their deep gratitude to
S.V. Peletminsky and to Yu.L. Bolotin for the attention paid to
this paper.

\bigskip

\begin{center}
\textbf{References}
\end{center}

[1] G.D. Birkhov, \textit{Dynamical Systems}, (American
Mathematical Society, Providence, Rode Island, 1927).

[2] N.S. Krylov, \textit{Works on the Foundation on Statistical
Physics}, (English translation: Princeton University Press,
Princeton, NJ, 1979).

[3] Ya.G. Sinai, Dynamical systems with elastic reflections,
\textit{Russ. Math. Serv}. \textbf{25}, 137 (1970); L.A.
Bunimovich, On ergodic properties of some billiards, Funct. Anal.
Appl. \textbf{8}, 73 (1974).

[4] Proc. of the Intern. Conf. on Classical and Quantum Billiards, \textit{J. Stat. Phys}.
\textbf{83}, No. 1-2 (1996).

[5] V.F. Lazutkin, \textit{KAM Theory and Semiclassical
Approximations to Eigenfunctions}, (Springer-Verlag,
Berlin-Heidelberg, 1993); M.G. Gutzwiller, \textit{Chaos in
Classical and Quantum mechanics}, (Springer-Verlag, New York,
1990).

[6] C.M. Marcus, A.J. Rimberg et al, Conductance fluctuations and
chaotic scattering in ballistic microstructures. \textit{Phys.
Rev. Lett}. \textbf{69}, 506 (1992); C. Ellegaard, T. Ghur et al,
Spectral statistics of acoustic resonances in aluminum blocs,
\textit{Phys. Rev. Lett}. \textbf{75}, 1546 (1995); H. Alt, H.D.
Graf et al, Chaotic dynamics in a three-dimensional
superconductiving microwave billiard, \textit{Phys.Rev. E}
\textbf{54}, 2303 (1996); J.U. Nocel, A.D. Stone, Ray and wave
chaos in assymmetric resonant optical cavities, \textit{Nature}
\textbf{385}, 45 (1997); and others.

[7] S.V. Naydenov, V.V. Yanovsky, The geometric-dynamic approach
to billiard systems. I.- II., \textit{Theor. and Math. Phys}.
\textbf{127}, \# 1, 500-512 (2001) [English edition];
\textbf{128}, 116 (2001) [Russian edition].

[8] V.I. Arnold, \textit{Dopolnitelnye Glavy Teorii Obyknovennych Differenzcialnych uravnenii},
(Moskva, Nauka, 1978) [in Russian].

[9] H.G. Shuster, \textit{Deterministic Chaos} (Springer, Heidelberg, 1982).

[10] A.I. Akhiezer, S.V. Peletminsky, \textit{Metody Statisticheskoy Fiziki},
(Moskva, Nauka, 1977) [in Russian].

\bigskip
\bigskip
\bigskip

\begin{center}
\textbf{Figure}
\end{center}

The geometry of a billiard and the schematic form of typical billiard
symmetric phase space. Elliptic zones of regular motion $R$, a chaotic
region $C$, the diagonal $\Delta $, lacunas $L$ and discriminants $D$.

\begin{figure}
\includegraphics[width=5.4 in,height=7.4in]{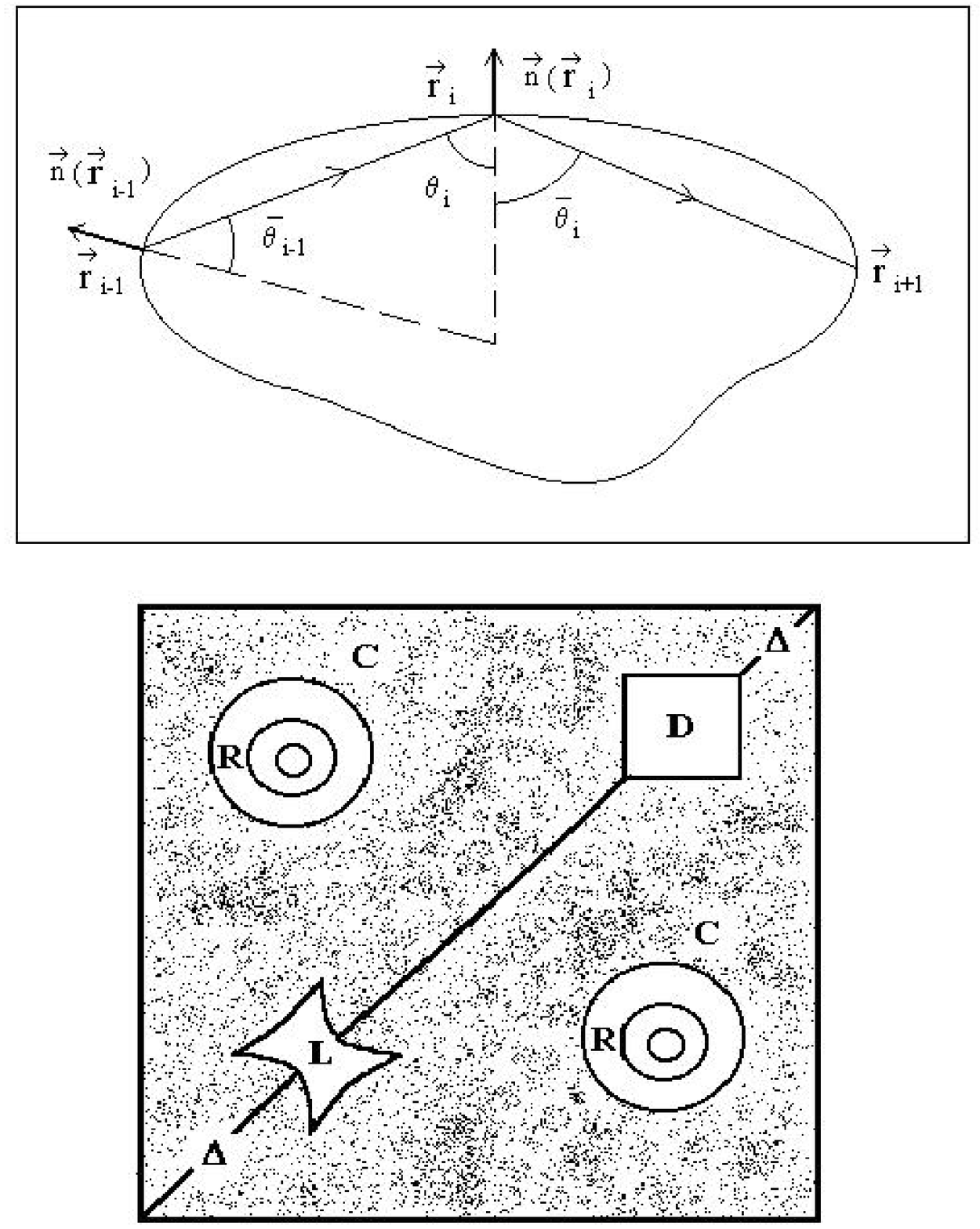}
\end{figure}

\end{document}